%% file: main.tex
\documentclass[sigconf,nonacm]{acmart}

\usepackage[english]{babel}
\usepackage[noend]{algpseudocode}
\usepackage[T1]{fontenc}
\usepackage[utf8]{inputenc}
\usepackage{algorithmicx}
\usepackage{algorithm}
\usepackage{amsmath}
\usepackage{amssymb}
\usepackage{caption}
\usepackage{enumitem}
\usepackage{float}
\usepackage{hyperref}
\usepackage{listings}
\usepackage{lstautogobble}
\usepackage{subcaption}
\usepackage{tikz}
\usepackage{todonotes}
\usepackage{color, colortbl}
\usepackage{multirow}
\usepackage{tikz-qtree}
\usepackage{tikz-qtree-compat}

\AtBeginDocument{

}

\definecolor{blue-cb}{RGB}{0,119,187}
\definecolor{cyan-cb}{RGB}{51,187,238}
\definecolor{teal-cb}{RGB}{0,153,136}
\definecolor{green-cb}{RGB}{34,136,51}
\definecolor{yellow-cb}{RGB}{170,156,49}
\definecolor{orange-cb}{RGB}{238,119,51}
\definecolor{red-cb}{RGB}{238,102,119}
\definecolor{magenta-cb}{RGB}{238,51,119}
\definecolor{purple-cb}{RGB}{170,51,119}
\definecolor{gray-cb}{RGB}{187,187,187}

\lstdefinelanguage{sql}{
    sensitive=false,
    keywords={select, from, where, as, in, and, distinct, create, drop, materialized, view, group, by, order, array_agg},
    keywordstyle=\color{green-cb},
    morekeywords=[2]{resultdb},
    keywordstyle = {[2]\color{red-cb}},
    morekeywords=[3]{begin, transaction, commit},
    keywordstyle = {[3]\color{purple-cb}},
    string=[b]',
    morestring=[b]",
    stringstyle=\color{orange-cb},
    comment=[l]{--},
    morecomment=[s]{/*}{*/},
    commentstyle=\color{blue-cb},
}

\newfloat{algfloat}{htbp}{lop}
\floatname{algfloat}{Algorithm}

\algrenewcommand{\algorithmiccomment}[1]{\hfill\textcolor{blue-cb}{$\triangleright$ \textit{#1}}}

\usetikzlibrary{
    arrows,
    calc,
    chains,
    matrix,
    positioning,
    scopes,
}

\begin{document}
\pagestyle{plain}
\title{What if an SQL Statement Returned a Database?}

\author{Joris Nix}
\affiliation{%
  \institution{Saarland University\\Saarland Informatics Campus}
}
\email{joris.nix@bigdata.uni-saarland.de}
\author{Jens Dittrich}
\affiliation{%
  \institution{Saarland University\\Saarland Informatics Campus}
}
\email{jens.dittrich@bigdata.uni-saarland.de}

\begin{abstract}
Every SQL statement is limited to return a single, possibly denormalized, table.
This design decision has far reaching consequences. (1.)~for databases users in terms of slow query performance, long query result transfer times, usability-issues of SQL in web applications and object-relational mappers.
In addition, (2.)~for database architects it has consequences when designing query optimizers leading to logical (algebraic) join enumeration effort, memory consumption for intermediate result materialization, and physical operator selection effort.
So basically, the entire query optimization stack is shaped by that design decision.

In this paper, we argue that the single-table limitation should be dropped.
We extend the SELECT-clause of SQL by a keyword `RESULTDB' to support returning a \textit{result database}.
Our approach has clear semantics, i.e.~our extended SQL returns subsets of all tables with only those tuples that would be part of the traditional (single-table) query result set, however \textit{without} performing any denormalization through joins. Our SQL-extension is downward compatible.

Moreover, we discuss the surprisingly long list of benefits of our approach.
First, for database users: far simpler and more readable application code, better query performance, smaller query results, better query result transfer times.
Second, for database architects, we present how to leverage existing closed source systems as well as change open source database systems to support our feature.
We propose a couple of algorithms to integrate our feature into both closed-source as well as open source database systems.
We present an initial experimental study with promising results.
\end{abstract}

\maketitle

\input{intro}
\input{return_database}
\input{rewrites}
\input{algorithm}
\input{related}
\input{eval}
\input{conclusion}

\begin{acks}
We thank Karl Bringmann for his algorithmic suggestions and feedback. We also thank Simon Rink for feedback and discussions on earlier versions of this paper.
\end{acks}

\balance
\bibliographystyle{ACM-Reference-Format}
\bibliography{main}

\end{document}

%% file: intro.tex

\section{Introduction}
\label{sec:introduction}
This paper is motivated by a real application.
Part of our group has recently started developing a web application.
When developing that application we ran into the following, on first sight seemingly easy, lecture-style problem.

\subsection{Motivating Problem and Running Example}
Consider a university application allowing users to retrieve information about professors and the lectures they give. The application may use any suitable UI-technology. The goal is to display information to the user as shown in~\autoref{fig:nested}.
\begin{figure}[h!]
    \centering
    \begin{small}
        \begin{tabular}{|l!{\vrule width 1.5pt}l|l|}
            \hline
            \rowcolor{black!40}
            \textbf{Professors} & \multicolumn{2}{c|}{{\textbf{Lectures}}} \\
            \hline
            \rowcolor{lightgray}
            name & name & difficulty \\
            \noalign{\hrule height 1.5pt}
            \multirow{2}{*}{Prof. A} & Computer Science & low \\
            \cline{2-3}
            & Databases & low \\
            \hline
            \multirow{2}{*}{Prof. B} & Computer Science & low \\
            \cline{2-3}
            & Databases & low \\
            \hline
        \end{tabular}
    \end{small}
    \caption{Desired result to be displayed by a UI as a relational structure.}
    \label{fig:nested}
\end{figure}

This is a nested, hierarchical display where each name of a professor is related to a set of lectures that professor gives.
Notice the many-to-many-relationship between professors and lectures.
In the example, both professors give both lectures `Computed Science' and `Databases'.
Also note that if we swapped columns `professors' and `lectures', the lecture data would not be repeated anymore, however the names of the professors would be repeated. If any data item is referenced multiple times it will be repeated in the display. Though it does not have to be a duplicate in the underlying data.

\subsection{Current Solution and its Problems}
\label{subsec:current-solutions}
Ok, we want to populate this hierarchical structure with data. Let's query the database.
We keep a relational database storing information about \emph{professors}, the \emph{lectures}, and information about which professors \emph{give} which lectures.
The excerpt of the entity-relationship model in~\autoref{fig:entity-relationship-model} represents this \textbf{many-to-many} relationship and will serve as a running example throughout the paper.
The corresponding relational model and sample data is shown in~Figure~\ref{tab:relations} (ignore the gray background for the moment).

At some point in the application stack, we generate and send the SQL statement shown in Figure~\ref{lst:example-query}.
That query returns the flat result table shown in~\autoref{tab:query-result}.
That flat result table can be used by the application to populate the hierarchical UI-component of~\autoref{fig:nested} with data.

At least in theory, actually, no: neither in theory nor in practice. The flat result table cannot be used directly as it has a number of problems that make it unsuitable for our purpose.

\begin{figure*}
    \centering
    \begin{subfigure}[h]{0.4\textwidth}
        \centering
        \usetikzlibrary{er}
        \begin{tikzpicture}[
            text depth=1pt,
            every attribute/.style={fill=black!0,draw=black},
            every entity/.style={fill=blue!20,draw=blue,thick},
            every relationship/.style={fill=orange!20,draw=orange,thick,aspect=1.5},
            scale = .8
        ]
            \node[entity] (professors)  at (0,0)   {Professors}
                child {node  [key attribute] {\underline{id}}}
                child {node  [attribute] {name}}
                child {node  [attribute] {age}};
            \node[entity, node distance=2.5cm] (lectures) [right=of professors] {Lectures}
                [sibling distance=2.2cm]
                child {node  [key attribute] {\underline{id}}}
                child {node  [attribute] {name}}
                child {node  [attribute] {difficulty}};
            \node[relationship, node distance=1em] [above=of $(professors.north)!0.5!(lectures.north)$] {give}
                edge (professors)
                edge (lectures);
            \node (N) at (0.9,0.8) {N};
            \node[right = 2.2cm of N] (M) {M};
        \end{tikzpicture}
        \caption{Entity-relationship model.}
        \label{fig:entity-relationship-model}
        \vspace*{0.4cm}
    \end{subfigure}
    \hfill
    \begin{subfigure}[h]{0.48\textwidth}
        \centering
        \begin{small}
        \hspace*{-1.2cm}
            \begin{minipage}{0.35\linewidth}
                \begin{tabular}{|c|c|c|}
                    \hline
                    \multicolumn{3}{|c|}{{\textbf{Professors}}} \\\hline\hline
                    \emph{\underline{id}} & \emph{name} & \emph{age} \\\hline
                    \rowcolor{lightgray}
                    0 & Prof. A & 49 \\
                    \rowcolor{lightgray}
                    1 & Prof. B & 60 \\
                    2 & Prof. C & 32 \\\hline
                \end{tabular}
                \newline
                \newline
                \newline
            \end{minipage}%
            \begin{minipage}{0.04\linewidth}
            ~
            \end{minipage}%
            \begin{minipage}{0.25\linewidth}
            \vspace*{1.3em}
                \begin{tabular}{|c|c|}
                    \hline
                    \multicolumn{2}{|c|}{{\textbf{give}}} \\\hline\hline
                    \emph{\underline{pid}} & \emph{\underline{lid}} \\\hline
                    \rowcolor{lightgray}
                    0 & 0 \\
                    \rowcolor{lightgray}
                    1 & 0 \\
                    \rowcolor{lightgray}
                    0 & 1 \\
                    \rowcolor{lightgray}
                    1 & 1 \\
                    0 & 2 \\
                    1 & 3 \\\hline
                \end{tabular}
            \end{minipage}%
            \begin{minipage}{0.35\linewidth}
                \begin{tabular}{|c|c|p{0.85cm}|}
                    \hline
                    \multicolumn{3}{|c|}{{\textbf{Lectures}}} \\\hline\hline
                    \emph{\underline{id}} & \emph{name} & \emph{difficulty} \\\hline
                    \rowcolor{lightgray}
                    0 & Computer Science &  low\\
                    \rowcolor{lightgray}
                    1 & Databases & low\\
                    2 & Mathematics &  high\\
                    3 & Artificial Intelligence  & high\\\hline
                \end{tabular}
                \newline
                \newline
            \end{minipage}%
            \vspace*{0.5em}
            \caption{Relational model and sample data.}
            \label{tab:relations}
        \vspace*{0.4cm}
        \end{small}
    \end{subfigure}
    \hfill
    \begin{subfigure}[h]{0.45\textwidth}
        \centering
        \begin{lstlisting}[language=sql]
        SELECT p.name, l.name, l.difficulty
        FROM professors AS p, give AS g, lectures AS l
        WHERE l.difficulty = 'low' AND
              p.id = g.pid AND
              l.id = g.lid
        ORDER BY p.name;
        \end{lstlisting}
        \caption{Q1: SQL statement.}
        \label{lst:example-query}
        \vspace*{0.4cm}
    \end{subfigure}
    \hfill
    \begin{subfigure}[h]{0.5\textwidth}
        \centering
        \definecolor{test}{rgb}{0.4,0.4,0.6}
        \begin{tabular}{|l|l|l|l|}
            \hline
            \multicolumn{3}{|c|}{{\textbf{Professors $\bowtie$ give $\bowtie$ Lectures}}} \\\hline\hline
            \emph{p.name} & \emph{l.name} & \emph{l.difficulty} \\\hline
            \colorbox{cyan-cb!70}{Prof. A} & \colorbox{magenta-cb!70}{Computer Science} & \colorbox{magenta-cb!70}{low} \\
            \colorbox{cyan-cb!70}{Prof. A} & \colorbox{orange-cb!70}{Databases} & \colorbox{orange-cb!70}{low} \\
            \colorbox{green-cb!70}{Prof. B} & \colorbox{magenta-cb!70}{Computer Science} & \colorbox{magenta-cb!70}{low} \\
            \colorbox{green-cb!70}{Prof. B} & \colorbox{orange-cb!70}{Databases} & \colorbox{orange-cb!70}{low} \\\hline
        \end{tabular}
        \caption{Relational result table. The colors represent attribute values of database entities (tuples) that get duplicated in the query result.
        }
        \label{tab:query-result}
    \end{subfigure}
    \caption{Running Example.}
    \label{fig:running-example}
\end{figure*}

To start, the result table contains redundancies, i.e.~the professor's names and lecture's names appear multiple times ($\rightarrow$\textbf{Problem~1: relational information redundancies}). It is not only that they are displayed multiple times, they are physically created multiple times. This is due to the underlying join denormalizing the data.
In ~\autoref{tab:query-result}, the color coding visualizes that effect. For instance, \colorbox{orange-cb!70}{Databases}, \colorbox{orange-cb!70}{low} appears twice.

In addition, the information that both values originate from the same tuple got completely lost in the result table. In theory, there could be two lectures named `Databases' and we would not be able to distinguish those in the result table without additionally projecting to their primary keys.
 ($\rightarrow$\textbf{Problem~2: relational information loss}).

In general, the larger the result table, i.e.~the more tuples in that table, the higher the likelihood to see these duplicates, the higher the effort to transmit these result sets to the client using the database ($\rightarrow$\textbf{Problem~3: data transmission costs}).

This is not so much a problem of analytical queries that perform grouping and aggregation, it is however an issue when the result set of a query is large.
In fact, as we will see later on, this may also be an issue for memory consumption while processing the query as those redundancies have to be materialized at some point in the physical execution plan. ($\rightarrow$\textbf{Problem~4: high memory requirements}).

What is also irritating about the result table is the fact that the notion of a `key' is not defined anymore, that notion is simply dropped when processing data in SQL or relational algebra
($\rightarrow$\textbf{Problem~5: key information loss}).

To remedy this, we could try to make the result table comply with Boyce-Codd-normal form~(BCNF). However, in the running example the keys of the base relations are not part of our result table.
Hence, we cannot argue along functional dependencies to verify whether the table is in a certain normal form.
As a workaround, we could make the specific tuples returned \textit{for this specific database instance} comply with BCNF. This can be done by considering the attributes \texttt{(p.name,l.name)} to form a composed key, i.e.~we define a schema for the result table as [\underline{pname:str, lname:str}, ldifficulty:str].
This approach breaks as soon as a second professor with the same name \textit{or} a second lecture with the same name got inserted into the underlying relations. This is precisely the reason why we teach undergraduates to never `normalize' any table against a specific database instance.

Back to the result table, SQL discards even more information from the underlying relations: where did a specific attribute in the result table come from in the first place? Does that attribute value correspond to an attribute in one of the base relations? Or was it computed? ($\rightarrow$\textbf{Problem~6: schema information loss}).
This again emphasizes that both SQL and relational algebra are not so much data \textit{retrieval} but in fact data \textit{transformation} languages: both languages take a set of base relations as their input and transform them into \textbf{a~single~output~relation}. However, considerable information on how the schema, keys, and data from the base relations are related to the schema, (possible) keys, and data in the single output relation gets lost on the way.

So, how do we use the flat result table (\autoref{tab:query-result}) to fill the UI-component (\autoref{fig:nested}) with data?
There are multiple options:

\noindent\textbf{Option~1:} We change the SQL statement to also return the keys, like that we would be able to write some code in our application to group the data by p.name and/or p.id.
In other words, our application will perform data processing.
Obviously, this feels weird, as of course the very idea of a database system is to provide data processing functionality such that application developers do not have to write data processing code over and over again.

\noindent\textbf{Option~2:} We do the same thing as in Option~1 but perform the grouping directly on the server using group-by, cube or any other suitable operator.
Then, we do however face the problem that we only want to group (in the sense of SQL's partition by) the data \textit{without} aggregating its values.
So we have to switch from group-by to using SQL's array-logic to return nested tables.
For instance, we can achieve this in SQL (using PostgreSQL-syntax) as shown in \autoref{lst:array-agg}.
\begin{lstlisting}[language=sql, caption={Modified query Q1 using SQL's array-logic.}, label={lst:array-agg}]
SELECT p.name,
       ARRAY_AGG (l.name || ':' || l.difficulty)
           AS nested
FROM professors AS p, give AS g, lectures AS l
WHERE l.difficulty = 'low' AND
      p.id = g.pid AND
      l.id = g.lid
GROUP BY p.name
ORDER BY p.name;
\end{lstlisting}

\autoref{fig:array-agg-result} shows the corresponding output. Note that \texttt{ARRAY\_AGG} creates a text[] column as specified in the ()-brackets.

\begin{figure}[h]
    \begin{tabular}{|l|l|}
        \hline
        \multicolumn{2}{|c|}{{\textbf{Professors $\bowtie$ give $\bowtie$ Lectures}}} \\\hline\hline
        \emph{p.name} & \emph{nested:text} \\\hline
        Prof. A & \{'Computer Science':'low','Databases':'low' \} \\
        Prof. B & \{'Computer Science':'low','Databases':'low' \} \\\hline
    \end{tabular}
    \caption{Output for the modified query Q1 given in \autoref{lst:array-agg}.}
    \label{fig:array-agg-result}
\end{figure}

Also be aware that if one of the inputs to \texttt{ARRAY\_AGG} is NULL, the entire output will be NULL and thus \textbf{none} of the non-NULL attribute values will be shown in the output! So extra logic is required to avoid these cases.
Good luck with that one!
In addition, the elements in an array must have the same type.
If that is not the case the source types must be cast into a common type which will in most cases be a string type.
So, again, in Option~2 we leave the relational world by transforming relational input data into arrays (or JSON or XML or objects: you name it), discarding all schema and relational information on the way.

In any way we achieve our goal, we have to spend some effort here to shoehorn the relational data from the database into our nested UI-component. ($\rightarrow$\textbf{Problem~7: data normalization effort}).

To avoid this problem, we could go for another approach:

\noindent\textbf{Option~3:} We send multiple queries to the database in the first place: one retrieving the set of qualifying professors, one retrieving the set of qualifying lectures, and one retrieving their relationships, i.e.~the relevant tuples from the `give' relation.
To be clear: it is not necessary to retrieve those three relations \textit{entirely}, we only need the subsets of tuples from those relations that are relevant to produce the query result of Q1, our original SQL statement in~\autoref{lst:example-query}.
In other words from the point of view of each relation we perform a semi-join with the rest of the query.

To not mess up isolation, we then wrap these three queries into a transaction as follows:

\begin{lstlisting}[language=sql, caption={Rewriting Q1 to return the relevant tuples from each relation only.}, label={lst:select-distinct-rewrite}]
BEGIN TRANSACTION;

SELECT DISTINCT p.id, p.name, p.age
-- Unmodified rest of query.

SELECT DISTINCT g.pid, g.lid
-- Unmodified rest of query.

SELECT DISTINCT l.id, l.name, l.difficulty
-- Unmodified rest of query.

COMMIT;
\end{lstlisting}

For our running example, this transaction produces three sets of tuples, i.e.~the ones marked with a gray background in~\autoref{tab:relations}. In the application code issuing this transaction, we have to make sure to assign the result sets of the three SQL statements to separate sets.

By retrieving these three sets we solve all of the problems mentioned so far: we do not create any redundant data in the result set (Problem~1). We do not lose any relational information (Problem~2).
We hopefully save some transmission costs by sending less data (Problem~3) and maybe even save some main memory while processing the query in the database system (Problem~4).
We do not lose any information about keys (Problem~5). And we do not lose any information about the underlying database schema (Problem~6). Finally, we do not have to spent effort to denormalize the data (Problem~7).

There is just one price we have to pay: ($\rightarrow$\textbf{New Problem: post-join effort}). On the client, we may have to \textit{post-join} the data, depending on what is required to feed the UI-components. However, as the join result displayed on a client is typically limited by what a user can see on a single screen anyhow, we expect this effort to be low.
That post-join effort was a no-go back in the 70ies when relational database technology was invented and users used dumb terminals. However, with the rise of powerful clients, this is not really a problem anymore except for extreme cases.

So, given this discussion, we were wondering, why not propose the following:

\subsection{Our Solution and our Contributions}

\textit{We extend SQL to allow it to return a result database, i.e.~only the tuples from those relations that are part of the query result.}

We will see that this not only makes the lives of application developers much easier but also has considerable impact on query optimization and processing.

In summary, the contributions of our paper are as follows:

\begin{enumerate}
\item We present a backward compatible SQL extension (SELECT RESULTDB) to allow SQL SELECT statements to return a clearly defined subset of a database rather than just a single table.
\item We present several rewrite algorithms allowing us to use any SQL-92-compliant closed source database system to support our extension.
\item We present an efficient native algorithm, allowing us to extend query optimizers to compute the result database efficiently directly inside a database system.
\item We present an initial experimental study comparing traditional single-table query processing with our rewrite methods.
    Our evaluation shows promising results to build upon.
\end{enumerate}

%% file: return_database.tex

\section{Querying a Database to return a Subdatabase}
\label{sec:return_database}

We propose to change SQL and relational algebra to return a \textit{subdatabase}. That subdatabase is well-defined: For each relation taking part in a query, we return the tuples that are part in forming the query result.

\subsection{Preliminaries}

Let $DB$ be a database containing a set of relations $\mathcal{R}=\{R_1,\ldots,R_n\}$.
Let $[R_i] =\{[ A_{i,1}:D_{i,1}, \ldots, A_{i,m(i)}: D_{i,m(i)}]\}, 1\leq i \leq n$ be the named schema of relation $R_i$ with attribute names $A_{i,j}$ and domains $D_{i,j}$ with $1\leq j\leq m(i)$.
The database schema, i.e.~the schema of $DB$, is the set of the schemas of the individual relations: $[DB] = \big\{[R_i] \;|\;\forall R_i \in \mathcal{R}\big\}$.

Let $Q_{\text{relation}}$ be a read-only query on $DB$, i.e.~$Q_{\text{relation}}$ is either a relational algebra expression or a SELECT statement\footnote{Actually, this definition also applies to similar query languages and also object-relational mappers and languages that are built on top of or are inspired by SQL, e.g.~the QuerySets language of Django ORM.}.
$Q_{\text{relation}}$ references a subset $\mathcal{R}' \subseteq \mathcal{R}$ of the relations in $DB$ as its input and returns a single relation $T$ as output:
\[
Q_{\text{relation}}\big(\mathcal{R}' \big) \mapsto T.
\]
Here $ T$ is a relation with a schema as defined in the SELECT-expression or final projection $\pi$ of $Q_{\text{relation}}$.

\subsection{A Query Returning a Subdatabase}

\begin{definition}[A Query Returning a Subdatabase]
\label{def:Q_DB}
Let $Q_{\text{DB}}$ be a query that returns a database as a result. It is defined as follows and relates to  $Q_{\text{relation}}$ as follows:
\begin{multline*}
Q_{\text{DB}}\big(\mathcal{R}' \big) \mapsto \mathcal{DB},\\
\mathcal{DB}:= \Big\{R_{i}.name\mapsto R_{i}' \;\big|\; R_{i} \in \mathcal{R}'\wedge R_{i}' := \pi_{[R_{i}]}\big( Q_{\text{relation}}(\mathcal{R}') \big) \subseteq  R_i  \Big\}.
\end{multline*}
\end{definition}
In other words, $Q_{\text{DB}}$ returns a \textit{subdatabase}. We return that subdatabase as a dictionary where for each relation $R_{i}$ in input relations $\mathcal{R}'$ maps the relation name to a subset of its tuples $R_{i}'\subseteq R_{i}$.
That subset is defined as the result of the original singe-table query $Q_{\text{relation}}$ projected to the schema of $R_i$, i.e.~the tuples of the different $R_{i}'$s that form the result set of $Q_{\text{relation}}$.
To avoid confusion here, the prefix `sub' in subdatabase refers to both (1.)~the fact that $Q_{\text{DB}}$ returns a subset of the relations in $DB$, and (2.)~that those relations are subsets of the relations in $DB$.

Notice the set semantics of relation algebra, i.e.~$\pi$ returns a duplicate-free set\footnote{Notice that the same semantics can be expressed in three different ways: (1.)~through a projection projection on a single relation, (2.)~through a group by without aggregation (which is an implicit distinct) grouping on attributes belonging to a single relation, or (3.)~through a rewritten query performing a semi-join of $R_i\ltimes Q_{\text{relation}\setminus R_i}(\mathcal{R}'\setminus R_i)$.}:
Tuples not being part of the result set are not returned.
In this definition, we assume that $Q_{\text{DB}}$ will return a database where each relation contains \textit{all} relations and \textit{all} their attributes as defined in~$\mathcal{R}' \subseteq \mathcal{R}$.

\subsection{Denormalizing Joins}
\label{sec:postjoin}
Given a query $Q_{\text{DB}}$, we can easily reconstruct the result of $Q_{\text{relation}}$ as follows:
\[
Q_{\text{relation}}\big(\mathcal{R}' \big)  = Q_{\text{relation}}\big(\mathcal{DB} \big) = Q_{\text{relation}}\Big(Q_{\text{DB}}\big(\mathcal{R}' \big) \Big).
\]
In other words, we can simply execute $Q_{\text{relation}}$ on the reduced database $\mathcal{R}$.

This implies, that we can always get back to the single-table result computed by $Q_{\text{relation}}$. It also implies that the computation of the single-table result can be done in two steps. i.e.~first, compute $Q_{\text{DB}}$ on a server; second, ship the resulting relations to the client; third, execute $Q_{\text{relation}}$ on the client. As a single, denormalized table as produced by $Q_{\text{relation}}$ could contain considerable redundancies, this might save bandwidth for shipping data to a client similar to standard semi-reducer joins in distributed query processing.

\subsection{Subschemas of Relations and Databases}
\label{subsec:subschemas}
So far, $Q_{\text{DB}}$ will return a subdatabase where each relation contains \textit{all} relations and \textit{all} their attributes as defined in $\mathcal{R}' \subseteq \mathcal{R}$.
However, we also want to be able to project data as in a standard SELECT-clause or the final (top-most) projection-operator $\pi$ in relational algebra.
To allow for this, we extend Definition~\ref{def:Q_DB} by projections:

\begin{definition}[Subschema of a Relation]
Let $R$ and $S$ be relations. $[R] \subseteq [S] \Leftrightarrow \big(\forall A \in [R]\Rightarrow\;A \in [S]\big)$. $[R]$ is called the \textit{relation subschema} of $[S]$.
\end{definition}

\begin{definition}[Subschema of a Database]
Let $[DB]$ and $[DB']$ be database schemas.
\begin{multline*}
[DB]' \subseteq [DB]  \Leftrightarrow  [DB]' = \big\{[R_i'] \subseteq [R_i]\forall R_i \in [DB'] \subseteq [DB]\big\}.
\end{multline*}
$[DB']$ is called the \textit{database subschema} of $[DB]$.
\end{definition}

\subsection{Projected Subdatabases}
\label{subsec:projections}

\begin{definition}[Projected Subdatabase]
Let  $Q_{\text{DB}}$ be a query returning a database $\mathcal{DB}$ with database schema $[\mathcal{DB}]$.
Then we can project $Q_{\text{DB}}$ to a database subschema $[\mathcal{DB}']$ as follows:
\[
\pi_{[\mathcal{DB}]'}\big(Q_{\text{DB}}(\mathcal{R'})\big) \mapsto \mathcal{DB'}
\]
The resulting \textit{projected subdatabase} is defined as:
\begin{align}
\mathcal{DB}':= \Big\{R_{i}.\text{name}\mapsto& R_{i}' \;|\; R_{i} \in \mathcal{DB}'\wedge& \label{eq:DB_projected}\\
&R_{i}' := \pi_{[R_{i}]\in [\mathcal{DB}']}\big( Q_{\text{relation}}(\mathcal{R}') \big) \subseteq  R_i  \Big\}. \nonumber
\end{align}
\end{definition}
So, in contrast to a subdatabase (\autoref{def:Q_DB}), a projected subdatabase may project to a subset of the relations referred to in a query. In addition, some attributes may be omitted from the result.
Again, the projection removes possible duplicate as in standard relational algebra projections.

\subsection{Relationship-Preserving Subdatabase Projection}
\label{subsec:relationship-preserving}

Let $[PK_{R_i}] \subseteq [R_i]$ with $[PK_{R_i}] \neq \emptyset$ be a non-empty subschema being the primary key of relation $R_i$.
Let
\begin{align*}
FK_{R_i} = \Big\{ [fk_j]  \;\big|\; [fk_j] \subseteq [R_i] \wedge \forall A\in [fk_j]&\\
 \text{ reference the same relation } R_j \in \mathcal{R}\Big\}
\end{align*}
be a (possibly empty) set of foreign key subschemas of $R_i$. Any foreign key subschema $[fk_j]$ may only refer one single relation in $[DB]$, including $R_i$ itself, i.e.~in the case of a recursive relationship.
  Let $\mathcal{R}_{[fk_j]}\in\mathcal{R}$ be the relation referred to by foreign key $[fk_j]$.

\begin{definition}[Relationship-Preserving Subdatabase Projection]
A~database projection $\pi_{[\mathcal{DB}]'}\big(\mathcal{DB}\big)$
 is called relationship-preserving iff:
\begin{align*}
 \forall_{1\leq i \leq n}\;\forall_{[fk_j] \in FK_{R_i}}:&\;\mathcal{R}_{[fk_j]} \in \mathcal{DB}' \\
&\; \wedge [PK_{\mathcal{R}_{[fk_j]}}] \subseteq [\mathcal{DB}]'\\
&\; \wedge [fk_j] \subseteq [\mathcal{DB}]'.
\end{align*}
\end{definition}
In other words, if the relations and their primary keys being referred to by those foreign keys as well as the referencing foreign keys are also contained in $[\mathcal{DB}']$, we call $[\mathcal{DB}']$ a relationship-preserving subdatabase projection.

If a post-join is to be performed along foreign-key constraints only (which we assume to be the 95\%-case), a subdatabase projection must be relationship-preserving.
If a post-join uses join predicates other than the foreign key constraints specified in the database schema, those attributes need to be included into the subdatabase projection as well.

\subsection{Extending SQL: \textsc{SELECT RESULTDB}}

As mentioned above, the idea to make a a query return a subdatabase is not restricted to one particular language, but can easily be added to many existing query languages. In this section, we briefly show how such an extension could look like in SQL.

We propose to extend the \textsc{SELECT} clause to
\[
\text{\textsc{SELECT RESULTDB }}[\mathcal{DB}']
\]
This statement will return $\pi_{[\mathcal{DB}]'}\big(Q_{\text{DB}}(\mathcal{R'})\big)$ as defined in~\autoref{subsec:projections}.
\autoref{lst:select-ind} shows a complete example with a SQL statement and the result set produced by that statement.

\begin{lstlisting}[language=sql, caption={Example SQL query showcasing our extension \texttt{SELECT RESULTDB} and the result dictionary.}, label={lst:select-ind}]
SELECT RESULTDB p.id, p.name,
                g.pid, g.lid,
                l.id, l.name, l.difficulty
FROM professors AS p, give AS g, lectures AS l
WHERE l.difficulty = 'low' AND
      p.id = g.pid AND
      l.id = g.lid;

Result = {
    'Professors': {
        (0, Prof. A),
        (1, Prof. B),
    },
    'give': {
        (0, 0),
        (1, 0),
        (0, 1),
        (1, 1),
    },
    'Lectures': {
        (0, Computer Science, low),
        (1, Databases, low),
    },
}
\end{lstlisting}

Notice the set semantics which avoids the confusion with SQL's multi-set semantic.
Also notice that \textsc{SELECT RESULTDB} does \textbf{not} imply that the contents of the subrelations returned have to be fully materialized: it is still possible to return results through pipelines as in standard single-table query processing. Think of it as each relation mapping to an iterator (aka cursor or any other suitable pull- or push-based streaming interface), see \autoref{lst:select-ind-stream}.

\begin{lstlisting}[language=sql, caption={Example for a stream-based result dictionary.}, label={lst:select-ind-stream}]
Result = {
    'Professors': stream_Professors,
    'give': stream_give,
    'Lectures': stream_Lectures,
}
\end{lstlisting}

Importantly, the user is not restricted in the choice of which attributes to return (as discussed in Sections~\ref{subsec:subschemas} and \ref{subsec:projections}).
Therefore, if the user is only interested in the professor and lecture names for example, only those two attributes would be returned.
In particular, the \textit{give} relation would not be part of the result database.
However, in order to be able to perform a \textit{post-join} on the client-side (as described in Section~\ref{sec:postjoin}), the database subschema has to be relationship-preserving (as described in Section~\ref{subsec:relationship-preserving}).

%% file: rewrites.tex

\section{SQL-based Rewrite Methods}
\label{sec:rewrite}

In this section, we discuss several SQL-based rewrite methods (RWM) that allow us to implement the query semantics introduced in~\autoref{sec:return_database}.
Since these rewrites happen entirely on the SQL level and do not require access to the source code of the database management system, this allows for very easy integration in every system, in particular closed source database systems.

Our rewrite methods can be classified along two dimensions:
\begin{enumerate}[wide, labelwidth=!, labelindent=0pt]
\item []\textbf{Dimension 1: SQL semi-join strategy.}
Unfortunately, SQL does not provide a keyword for semi-joins.
Hence we have to write semi-joins implicitly.
This can be done in two ways: either through an inner join followed by a \textsc{SELECT DISTINCT} on a single relation, see~\autoref{lst:select-distinct-rewrite}.
Alternatively, we can use a subquery with \textsc{IN}, see~\autoref{lst:subquery-rewrite}. Semantically, both rewrites express the same query and return the same result set. Unfortunately, the two rewrites may yield very different plans in many database systems. Hence, each rewrite strategy can often be considered an implicit hint to the optimizer (breaking the strict separation of SQL's declarativeness from query optimization). Due to these effects,  we will consider both rewrites for this paper.
\item []\textbf{Dimension~2: Materialization strategy.}
For the different rewrites, we can prematerialize or cache data, e.g.~by issuing suitable materialized views. These materialized views can precompute the entire query result (which is trivial) or suitable parts of the query which allow us to achieve a good trade-off between prematerialization effort and query processing time.
\end{enumerate}
In total, these two dimensions span a two-dimensional landscape, see~\autoref{fig:landscape}.
For each combination of semi-join and materialization strategy, we show how the rewrite methods that we are going to discuss in the following subsections can be categorized.

\begin{figure}[h!]
    \begin{tabular}{c!{\vrule width 1.5pt}c|c}
        & SELECT DISTINCT & Subquery \\
        \noalign{\hrule height 1.5pt}
        Dynamic & RWM~1 & RWM~3 \\
        \hline
        Materialized & RWM~2 & RWM~4
    \end{tabular}
    \caption{SQL semi-join $\times$ materialization landscape.}
    \label{fig:landscape}
\end{figure}

\subsection{Rewrite Method~1: Dynamic SELECT DISTINCT}
\label{subsec:select-distinct}
The first rewrite method transforms the \textit{single} \textsc{SELECT RESULTDB} query into \textit{multiple} \textsc{SELECT DISTINCT} queries, one for each relation that participates in the output.
Each of those queries computes the same join result internally. Then, each query projects to the unique attributes of exactly \textit{one} relation.
Therefore, we need as many \textsc{SELECT DISTINCT} queries as we have different relations in the \textsc{SELECT} clause.
Furthermore, we use the \textsc{DISTINCT} keyword to ensure that no duplicated entities due to the join are returned.
Note, that we can also omit this keyword if the primary key is part of the \textsc{SELECT} clause.
We discussed this option already in the introduction as Option~3, see~\autoref{lst:select-distinct-rewrite}.

The advantage of this rewrite method is that it is relatively straightforward and easy to implement. We only have to check if the query contains the \textsc{RESULTDB} keyword and if so, construct the different queries and send them to the database inside a transaction.
However, the most apparent disadvantage is that essentially the same query with a slightly different \textsc{SELECT} clause gets executed multiple times and with that, a potentially expensive join. We can only hope that the database systems underneath recognizes that the transaction contains similar queries and therefore, is able to cache the query execution plan or the join result.

\subsection{Rewrite Method~2: Materialized SELECT DISTINCT}
\label{subsec:materialized-view}
To proactively circumvent that a potentially expensive join is executed multiple times is likely the case for the rewrite method presented in~\autoref{subsec:select-distinct}. Therefore, we introduce another rewrite method that makes use of \textsc{MATERIALIZED VIEW}s (MVs).
In this rewrite method, first, we explicitly create a (temporary and unmaintained) materialized view \textit{once}. This represents a snapshot as of the time the MV is created. Hence, there is no need to wrap queries into a transaction as in Rewrite~Method~1. Second, we execute the individual queries against that MV. Third, we drop the MV.
\autoref{lst:mat-view-rewrite} shows the corresponding queries.
\begin{lstlisting}[language=sql, caption={Rewrite of the \texttt{SELECT RESULTDB} query using a \texttt{MATERIALIZED VIEW} and multiple \texttt{SELECT DISTINCT}s.}, label={lst:mat-view-rewrite}]
CREATE MATERIALIZED VIEW MV AS
SELECT p.id, p.name,
       g.pid, g.lid,
       l.id, l.name, l.difficulty
FROM professors AS p, give AS g, lectures AS l
WHERE l.difficulty = 'low' AND
      p.id = g.pid AND
      l.id = g.lid;

SELECT DISTINCT p.id, p.name
FROM MV;

SELECT DISTINCT g.pid, g.lid
FROM MV;

SELECT DISTINCT l.id, l.name, l.difficulty
FROM MV;
\end{lstlisting}

If the database system does not support MVs, a similar rewrite strategy can be achieved by creating a new temporary table \textit{foo} and selecting into it using \textsc{INSERT INTO} \textit{Foo} \textsc{SELECT ...}.
An even more technical approach which is only available in certain DBMS would be to use a result cache like Oracle's RESULT\_CACHE hint\footnote{\url{https://docs.oracle.com/en/database/oracle/oracle-database/19/tgdba/tuning-result-cache.html}}.

In either way, Rewrite Method~2 has the advantage that we avoid the potentially expensive computation of the same query over and over again. However, the disadvantage is that we have to materialize a potentially large query result inside the DBMS. This may be especially costly in a disk-based DBMS.

Overall, Rewrite Method~2 trades upfront costs for materializing the join result versus repeated cost for the computation of the join result. Therefore, depending on the size of the join result and the number of queries making use of the MV, Rewrite Method~2 could yield a better performance in terms of overall execution time.

\subsection{Rewrite Method~3: Dynamic Subquery}
Unfortunately, we have had the experience that some database systems are unable to identify the underlying semi-join semantic of \textsc{SELECT DISTINCT} queries.
Therefore, we have to steer the query optimizer into using semi-joins internally by providing a hint using subquery syntax.
\autoref{lst:subquery-rewrite} shows the different dynamic subqueries, i.e.\ without prematerializing any join result.
We explicitly include the \textsc{DISTINCT} keyword here to ensure a duplicate-free result since we do not necessarily need to project to the primary keys.

\begin{lstlisting}[language=sql, caption={Rewrite of the \texttt{SELECT RESULTDB} query using different dynamic subqueries.}, label={lst:subquery-rewrite}]
BEGIN TRANSACTION;

SELECT DISTINCT p.id, p.name
FROM professors AS p
WHERE p.id IN -- p.id = g.pid
    (SELECT g.pid
     FROM give AS g, lectures AS l
     WHERE l.difficulty = 'low' AND
           l.id = g.lid);

SELECT DISTINCT g.pid, g.lid
FROM give AS g
WHERE g.pid IN -- g.pid = p.id
    (SELECT p.id
     FROM professors AS p) AND
      g.lid IN -- g.lid = l.id
    (SELECT l.id
     FROM lectures AS l
     WHERE l.difficulty = 'low');

SELECT DISTINCT l.id, l.name, l.difficulty
FROM lectures AS l
WHERE l.difficulty = 'low' AND
      l.id IN -- l.id = g.lid
        (SELECT g.lid
         FROM professors AS p, give AS g
         WHERE p.id = g.pid);

COMMIT;
\end{lstlisting}
Note that for the \textit{professors} and \textit{lectures} relations we only need one subquery since they both are associated with exactly one join predicate, while the \textit{give} relation requires two subqueries as it has two foreign keys.

Using this rewrite, the query optimizer is in some cases able to efficient execute a semi-join instead of an inner join that potentially produces duplicates that we need to eliminate again afterwards anyways.
In addition, there is no need to materialize a potentially large join result.
However, in comparison to the other rewrite methods, this rewrite is arguably a little more involved.
Furthermore, whether the query optimizer is really able use a semi-join operator depends on many factors\footnote{\url{https://dev.mysql.com/doc/refman/8.0/en/semijoins.html}}.
Therefore, if the underlying query is somewhat more complex (e.g.\ contains a \textsc{HAVING} clause or any aggregate function), the query optimizer will likely be unable to apply this optimization.

\subsection{Rewrite Method~4: Materialized Subquery}

Actually, for Materialized Subqueries we have several options.
We just focus on the most promising option here. If we inspect~\autoref{lst:subquery-rewrite}, we notice that each of the three subqueries is different.
So, imagine we created a separate materialized view for each of those subqueries.
Then, each of those MVs would only be used once by the actual query.
Hence, w.r.t.~the sum of both times, i.e.~first, compute the MV and, second, execute the actual query, we would not gain anything. Actually, we would probably even lose performance due to materializing an intermediate result that could be pipelined.
So, ideally we should seek to prematerialize something smaller that can then be reused by several of the semi-join queries.
One option is to compute a full join index, i.e.~for all participating relations, only project to their primary keys.
The additional foreign keys can be dropped as they are redundant. \autoref{lst:mat-view-rewrite-subquery} shows an example.

\begin{lstlisting}[language=sql, caption={Rewrite of the \texttt{SELECT RESULTDB} query using a \texttt{MATERIALIZED VIEW} and two subqueries.}, label={lst:mat-view-rewrite-subquery}]
CREATE MATERIALIZED VIEW MV AS
SELECT p.id, l.id,
FROM professors AS p, give AS g, lectures AS l
WHERE l.difficulty = 'low' AND
      p.id = g.pid AND
      l.id = g.lid;

SELECT DISTINCT p.id, p.name
FROM professors AS p
WHERE p.id IN (SELECT p.id from MV);

SELECT DISTINCT *
FROM MV;

SELECT DISTINCT l.id, l.name, l.difficulty
FROM lectures AS l
WHERE l.id IN (SELECT l.id from MV);
\end{lstlisting}

In~\autoref{lst:mat-view-rewrite-subquery}, Query~1 returns all distinct tuples from the professors relation whose \texttt{p.id}.is contained in the MV. Query~3 returns all distinct tuples from the lectures relation whose \texttt{l.id} is contained in the MV. Query~2 does not even have to compute anything but just returns the MV as is.

%% file: algorithm.tex

\section{SELECT RESULTDB Algorithm}
\label{sec:algorithm}
In~\autoref{sec:rewrite}, we focused on how we can implement queries returning a database only using SQL-based rewrites.
While this approach is widely applicable due to its ease of implementation and because we do not need access to the source code, each presented method comes with its own drawbacks.
We either perform a potentially expensive join multiple times, materialize a large join result, or try to enforce a semi-join optimization using subqueries.

Therefore, in this section, we present an algorithm that can be integrated into a DBMS to efficiently compute the result set of our \textsc{SELECT RESULTDB} queries.
In particular, our algorithm allows us to fully reduce all relations of a join graph with an arbitrary topology.
For this, we will first discuss how we can leverage Yannakakis' algorithm for the subproblem of acyclic join graph topologies in~\autoref{subsec:acyclic}.
Afterwards, in~\autoref{subsec:cyclic} we show how we can transform cyclic queries into acyclic ones and with that, reuse the algorithm from~\autoref{subsec:acyclic}.
Putting it all together, we present our final algorithm in ~\autoref{subsec:putting-it-all-together}.

\subsection{Preliminaries}
\label{subsec:alg-prelim}
The core idea is to efficiently \textit{reduce} each individual relation to the minimal set of tuples that participate in the result set as defined in \autoref{def:Q_DB}.
For this, we are going to make use of \textit{semi-joins}.

\begin{definition}[(Left) Semi-Join]
    Let $R$ and $S$ be two relations and $[R'] \subseteq [R]$, $[S'] \subseteq [S]$ with $|[R']| = |[S']|$ be two subschemas of the same length, then the (left) semi-join between $R$ and $S$ is defined as:
    $$ R \ltimes_{[R'],[S']} S = \pi_{[R]}(R \bowtie_{[R'],[S']} S)$$
    where $R \bowtie_{[R'],[S']} S$ is the equi-join between $R$ and $S$ based on the equality of the tuples defined by the subschemas $[R']$ and $[S']$.
\end{definition}
Note that in contrast to the join operator, the semi-join is not commutative. In the following, we will refer to a \textit{left} semi-join when we talk about semi-joins.
Furthermore, we omit the subschemas if it is clear from the join predicate along which attributes the semi-join is performed.

Semi-joins were already introduced by Edgar F. Codd in his seminal paper introducing the relational model~\cite{codd_relational-model}. However, Codd coined this operation \textit{restriction} at that time, which nicely expresses the underlying functionality.
Similar to this and as done in previous work~\cite{bernstein_semi-joins}, we will be using the term \textit{semi-join reduction} or just \textit{reduction} when performing a semi-join.
In particular, we say ``$R$ is reduced by $S$'' or ``$S$ reduces $R$'' if we perform $R \ltimes S$.
Depending on the context, a semi-join reduction can also refer to the reduction of every relation that is part of a query.

As already shown in previous work~\cite{bernstein_semi-joins}, the shape of the join graph is an essential factor for computing a semi-join reduction.
A join graph for some query $\mathcal{Q}$ is defined as $\mathcal{JG} = (\mathcal{R}, \mathcal{J})$, where $\mathcal{R}$ is the set of relations and $\mathcal{J}$ is the set of joins in $\mathcal{Q}$.

\subsection{Acyclic Join Graph Topology}
\label{subsec:acyclic}
The Yannakakis algorithm~\cite{yannakakis_acyclic-database-schemes} already provides an efficient way for solving acyclic (tree) conjunctive queries, i.e.~queries that correspond to select-project-join (SPJ) queries and that have a join graph without any cycles.
The algorithm essentially works in three main steps.
After choosing an arbitrary root node (every node in a tree can act as a root), we first perform consecutive semi-joins bottom-up from the \textit{leaves} to the \textit{root} node.
Second, we perform consecutive semi-joins top-down from the \textit{root} to the \textit{leaves}.
Lastly, the Yannakakis algorithm joins the reduced relations to obtain a single output relation.
The main motivation of this algorithm is to keep intermediate results as small as possible by reducing all relevant relations to their minimal set of tuples that participate in the join before actually joining the relations (and producing a potentially large join result).
\autoref{alg:yannakakis} shows the high level steps of this algorithm.

\begin{algorithm}[H]
    \centering
    \begin{algorithmic}[1]
        \Statex (0) Choose an arbitrary node in the join graph as root.
        \Statex (1) Perform bottom-up semi-joins from leaves to root.
        \Statex (2) Perform top-down semi-joins from root to leaves.
        \Statex (3) Compute join result.
    \end{algorithmic}
    \caption{Yannakakis Algorithm.}
    \label{alg:yannakakis}
\end{algorithm}

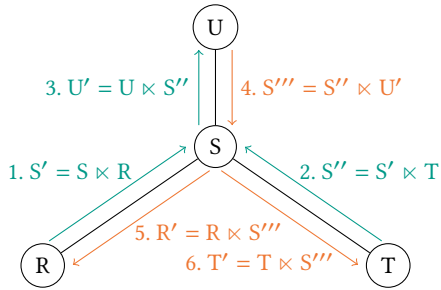
\begin{figure}[!h]
    \centering
    \begin{tikzpicture}[
        v/.style={circle, draw=black, minimum size=10pt}
    ]
            \node[v] (R) at (0,0) {R};
            \node[v, node distance=4cm] (T) [right=of R] {T};
            \node[v] (S) [above=of $(R.north)!0.5!(T.north)$] {S};
            \node[v] (U) [above=of S] {U};

            \draw (R) -- (S);
            \draw (S) -- (T);
            \draw (S) -- (U);
            \begin{scope}[transform canvas={xshift=-.5em, yshift=.5em}]
                \draw [->, draw=teal-cb] (R) -- (S);
                \node[left=of $(R.north)!0.5!(S.south)$, xshift=13mm, yshift=3mm] {\textcolor{teal-cb}{1. $\text{S}' = \text{S}\ltimes\text{R}$}};
            \end{scope}
            \begin{scope}[transform canvas={xshift=.5em, yshift=.5em}]
                \draw [->, draw=teal-cb] (T) -- (S);
                \node[right=of $(T.north)!0.5!(S.south)$, xshift=-13mm, yshift=3mm] {\textcolor{teal-cb}{2. $\text{S}'' = \text{S}'\ltimes\text{T}$}};
            \end{scope}
            \begin{scope}[transform canvas={xshift=-.7em}]
                \draw[->,draw=teal-cb] (S) -- (U);
                \node[left=of $(S.north)!0.5!(U.south)$, xshift=1cm] {\textcolor{teal-cb}{3. $\text{U}' = \text{U}\ltimes\text{S}''$}};
            \end{scope}
            \begin{scope}[transform canvas={xshift=.7em}]
                \draw[->,draw=orange-cb] (U) -- (S);
                \node[right=of $(U.south)!0.5!(S.north)$, xshift=-1cm] {\textcolor{orange-cb}{4. $\text{S}''' = \text{S}''\ltimes \text{U}'$}};
            \end{scope}
            \begin{scope}[transform canvas={xshift=.5em, yshift=-.5em}]
                \draw [->, draw=orange-cb] (S) -- (R);
                \node[right=of $(S.south)!0.5!(R.north)$, xshift=-12mm, yshift=-2mm] {\textcolor{orange-cb}{5. $\text{R}' = \text{R}\ltimes\text{S}'''$}};
            \end{scope}
            \begin{scope}[transform canvas={xshift=-.5em, yshift=-.5em}]
                \draw [->, draw=orange-cb] (S) -- (T);
                \node[left=of $(S.south)!0.5!(T.north)$, xshift=17mm, yshift=-6mm] {\textcolor{orange-cb}{6. $\text{T}' = \text{T}\ltimes\text{S}'''$}};
            \end{scope}
    \end{tikzpicture}
    \caption{Acyclic join graph topology visualizing a possible semi-join order with U as root. \textcolor{teal-cb}{$\rightarrow$} shows the bottom-up and \textcolor{orange-cb}{$\rightarrow$} the top-down pass.}
    \label{fig:acyclic-topology}
\end{figure}

\autoref{fig:acyclic-topology} shows an acyclic join graph consisting of four relations and visualizes \textit{one possible} semi-join order.
In this example, we choose the node U as root. Afterwards, we compute a breadth-first search order for the edges starting at the root.
The reversed order now gives us a suitable sequence for the bottom-up semi-joins while we perform the top-down semi-joins in the original order.
The semi-joins are performed in the direction of the arrows, i.e.~an arrow from R to S represents the semi-join S $\ltimes$ R.
Note, that for a specific node with multiple children, the order in which the semi-joins are applied does not matter.

\autoref{alg:select-resultdb} shows pseudo-code for the computation of our result database based on the fundamental steps in Yannakakis' algorithm depicted in \autoref{alg:yannakakis}.
In line~2, we choose an arbitrary root node of our tree-structured join graph (step 0).
In line~3, we order the edges of the join graph, i.e. the joins, in a breadth-first-search order starting at the root node.
Further, we assume that the join operands are ordered with respect to the root node, i.e. we essentially have directed edges from the root to the leaves.
This is important since the semi-join operation is not commutative and we have to ensure that the semi-joins are performed in the correct direction.
In lines~4-5, we perform the bottom-up semi-joins with some join predicate (step 1) and in lines~6-7, we perform the top-down semi-joins (step 2).
In contrast to Yannakakis' algorithm however, we do not have to execute step 3 anymore, i.e.~we do not have to compute the final join result.
The reason for this is, that the reduced relations already constitute our desired result database.
This is a very crucial characteristic of our \textsc{SELECT RESULTDB} queries.

The probably most apparent explanation why Yannakakis' algorithm is not used in traditional query processing is the uncertainty whether the overhead of the semi-join reduction can be compensated by smaller subsequent join results.
Algorithmically, a semi-join and a traditional equi-join perform the same steps.
Both algorithms build a hash table on one side and probe the other.
Therefore, performing these semi-joins beforehand on a single node system (distributed settings are a different matter) is most likely even more costly.
However, we do not have to worry about this as we are only interested in the reduced relations.
This allows us to fully utilize the efficiency of Yannakakis' algorithm while avoiding its potential disadvantages.

\begin{algorithm}[H]
    \centering
    \begin{algorithmic}[1]
        \Function{reduce\_relations}{$G$} \Comment{$G$ is an acyclic join graph}
            \State root = choose\_arbitrary\_node($G$) \Comment{(0) arbitrary root node}
            \State edges\_bfs\_order = bfs\_edges($G$, root)
            \For{join $\in$ reversed(edges\_bfs\_order)} \Comment{(1) bottom-up}
                \State semi\_join(join.right, join.left, join.pred)
            \EndFor
            \For{join $\in$ edges\_bfs\_order} \Comment{(2) top-down}
                \State semi\_join(join.left, join.right, join.pred)
            \EndFor
            \State return $G$ \Comment{$G$ contains reduced relations}
        \EndFunction
    \end{algorithmic}
    \caption{Reduce relations in join graph $G$ using Yannakakis' algorithm.}
    \label{alg:select-resultdb}
\end{algorithm}

\subsection{Cyclic Join Graph Topology}
\label{subsec:cyclic}
Yannakakis' algorithm is explicitly only defined for acyclic conjunctive queries.
Furthermore, Bernstein et al.~\cite{bernstein_semi-joins} show that for cyclic queries, semi-joins either cannot be used to \textit{fully reduce} the relations or we need a very long sequence of semi-joins.
Note that by ``fully reduce'' we refer to the state of a relation where only the minimal set of tuples that participate in the final join result remain.
In \autoref{fig:acyclic-topology} for example, after performing the first semi-join S~$\ltimes$~R, the relation S is only partially reduced.
Only after S is reduced by all neighboring relations (which in turn have to be fully reduced as well), the relation S cannot be further reduced anymore and thus, we call it fully reduced.

\autoref{fig:counterexample} gives a minimal example that showcases why Yannakakis' algorithm cannot be applied to cyclic queries.
Assume we have the data given in Figure~\ref{tab:cycle-data} and the join graph with a cyclic topology in Figure~\ref{fig:cycle-topology}.
The actual join result only consists of a single row depicted in Figure~\ref{tab:join-result-cycle}.
The corresponding result database should therefore also only contain exactly \textit{one} tuple in each result relation.
Figure~\ref{tab:select-ind-result-cycle} shows the expected result database.
Applying Yannakakis' algorithm on this example has two fundamental problems. First, we cannot even choose a root node for this join graph.
Second, no matter in which order we apply potential semi-joins, we will never filter out any of the tuples from our example data.
The reason is that every row in each relation has a join partner in every other relation.
As a result, Yannakakis' algorithm essentially produces the original subdatabase shown in Figure~\ref{tab:cycle-data} and we cannot obtain our expected output.

\begin{figure*}
    \centering
    \begin{subfigure}[h]{0.49\textwidth}
        \centering
        \begin{minipage}{0.3\linewidth}
            \centering
            \begin{tabular}{|c|}
                \hline
                \textbf{R} \\\hline\hline
                \emph{\underline{id}} \\\hline
                0 \\
                1 \\\hline
            \end{tabular}
        \end{minipage}%
        \begin{minipage}{0.3\linewidth}
            \centering
            \begin{tabular}{|c|c|}
                \hline
                \multicolumn{2}{|c|}{{\textbf{S}}} \\\hline\hline
                \emph{\underline{rid}} & \emph{\underline{tid}} \\\hline
                0 & 0 \\
                0 & 1 \\
                1 & 0 \\\hline
            \end{tabular}
        \end{minipage}%
        \begin{minipage}{0.3\linewidth}
            \centering
            \begin{tabular}{|c|}
                \hline
                \textbf{T} \\\hline\hline
                \emph{\underline{id}} \\\hline
                0 \\
                1 \\\hline
            \end{tabular}
        \end{minipage}%
        \caption{Example data for our minimal counterexample. This subdatabase also represents the output that would be produced using Yannakakis' algorithm.}
        \label{tab:cycle-data}
    \end{subfigure}
    \hfill
    \begin{subfigure}[h]{0.49\textwidth}
        \centering
        \begin{tikzpicture}[
            v/.style={circle, draw=black, minimum size=10pt}
        ]
                \node[v] (S) at (0,0) {S};
                \node[v] (R) [below left=of S] {R};
                \node[v] (T) [below right=of S] {T};

                \draw (R) -- (S) node[midway, sloped, above] {id=rid};
                \draw (S) -- (T) node[midway, sloped, above] {tid=id};
                \draw (R) -- (T) node[midway, sloped, above] {id=id};
        \end{tikzpicture}
        \caption{Cyclic join graph topology.}
        \label{fig:cycle-topology}
    \end{subfigure}
    \hfill
    \vspace{5mm}
    \begin{subfigure}[h]{0.49\textwidth}
        \centering
        \begin{tabular}{|c|c|c|c|}
            \hline
            \multicolumn{4}{|c|}{{\textbf{R$\bowtie$S$\bowtie$T}}} \\\hline\hline
            \emph{R.id} & \emph{S.rid} & \emph{S.tid} & \emph{T.id}\\\hline
            0 & 0 & 0 & 0\\\hline
        \end{tabular}
        \caption{Actual join result.}
        \label{tab:join-result-cycle}
    \end{subfigure}
    \hfill
    \begin{subfigure}[h]{0.49\textwidth}
        \centering
        \begin{minipage}{0.3\linewidth}
            \centering
            \begin{tabular}{|c|}
                \hline
                \textbf{R} \\\hline\hline
                \emph{\underline{id}} \\\hline
                0 \\\hline
            \end{tabular}
        \end{minipage}%
        \begin{minipage}{0.3\linewidth}
            \centering
            \begin{tabular}{|c|c|}
                \hline
                \multicolumn{2}{|c|}{{\textbf{S}}} \\\hline\hline
                \emph{\underline{rid}} & \emph{\underline{tid}} \\\hline
                0 & 0 \\\hline
            \end{tabular}
        \end{minipage}%
        \begin{minipage}{0.3\linewidth}
            \centering
            \begin{tabular}{|c|}
                \hline
                \textbf{T} \\\hline\hline
                \emph{\underline{id}} \\\hline
                0 \\\hline
            \end{tabular}
        \end{minipage}%
        \caption{Expected \texttt{SELECT RESULTDB} output.}
        \label{tab:select-ind-result-cycle}
    \end{subfigure}
    \caption{Counterexample that shows why Yannakakis' algorithm cannot be applied to cyclic queries.}
    \label{fig:counterexample}
\end{figure*}

So, in order to compute a semi-join reduction for cyclic queries, we essentially have two options.
First, we could try to come up with a new algorithm that works on cyclic queries as well.
Or second, instead of developing a completely new algorithm from scratch, we could somehow ``transform'' cyclic queries into tree queries and leverage Yannakakis' algorithm again.
We decided to pursue the second option.
The general idea is to \emph{fold} relations together such that the cycles in the join graph are resolved.
\autoref{alg:folding} shows pseudo-code for the construction of an acyclic join graph.
At the core, the algorithm consists of two steps. First, we choose a (random) node $x$ in the join graph and one of its neighbors $y$ (lines~3-4).
Second, we replace those two nodes $x$ and $y$ with the join result of both relations in the join graph and adjust other affected joins accordingly (line~5).
We repeat these steps until our join graph is acyclic (line~2).
Since our join graph is undirected and connected, we can check if it is cyclic very easily by comparing the number of joins and the number of relations.
If we have at least as many joins as relations, we inevitably have a cyclic join graph.

\begin{algorithm}[H]
    \centering
    \begin{algorithmic}[1]
        \Function{fold\_join\_graph}{$G$}
            \While{$G$.is\_cyclic()} \Comment{\#joins >= \#relations}
                \State $x$ = choose\_node($G$)
                \State $y$ = choose\_neighbor($G$, $x$)
                \State $G$.replace($x$, $y$, $x \bowtie y$) \Comment{replace relations x and y with the join of both relations; adjust affected joins accordingly}
            \EndWhile
            \State return $G$ \Comment{acyclic join graph}
        \EndFunction
    \end{algorithmic}
    \caption{Transform cyclic join graph into acyclic one.}
    \label{alg:folding}
\end{algorithm}
Note, that we do not specify how to choose either node $x$ or $y$.
We can choose them arbitrarily.
However, we are most likely interested in a folding of the join graph in an acyclic one that only requires the minimal number of joins as well as the set of joins with overall minimal computation cost.
We coin this problem the \emph{Tree Folding Enumeration Problem}.
Exploring this problem goes beyond the scope of this paper and we will investigate it as part of future work.

\autoref{fig:folding} shows an example transformation of a cyclic join graph into an acyclic one by folding multiple vertices.
Our initial join graph (JG~1) has multiple cycles (e.g.\ R\textcolor{blue-cb}{--}S\textcolor{yellow-cb}{--}U\textcolor{purple-cb}{--}T\textcolor{green-cb}{--}R or R\textcolor{blue-cb}{--}S\textcolor{orange-cb}{--}T\textcolor{green-cb}{--}R).
In the first step, we arbitrarily choose the nodes T and U and replace them in the join graph with their join result T$\bowtie$U.
Note that we now have a conjunctive join predicate between S and T$\bowtie$U which is visualized by having multiple edges in their respective color between two nodes.
After this first transformation, we still have a cyclic join graph at hands (JG~2).
Therefore, we repeat the process and join the arbitrarily chosen nodes R and S.
The final join graph (JG~3) consists of two nodes connected by essentially three join predicates.

Note, that this is just \emph{one} possible outcome.
If we had chosen the nodes S and T in the first step, we would have obtained an acyclic join graph already after folding the first two nodes.

\begin{figure}[h]
    \centering
    \begin{subfigure}[t]{0.3\columnwidth}
        \centering
        \begin{tikzpicture}[
            v/.style={circle, draw=black, minimum size=26pt},
            e/.style={ultra thick}
        ]
                \node[v] (R) at (0,0) {R};
                \node[v] (S) [right=of R] {S};
                \node[v] (T) [below=of R] {T};
                \node[v] (U) [below=of S] {U};

                \draw[e, blue-cb] (R) to (S);
                \draw[e, green-cb] (R) to (T);
                \draw[e, yellow-cb] (S) to (U);
                \draw[e, orange-cb] (S) to (T);
                \draw[e, purple-cb] (T) to (U);
        \end{tikzpicture}
        \caption{JG~1}
        \label{fig:jg1}
    \end{subfigure}
    \begin{tikzpicture}
        \node (og) at (0,0) {};
        \node (T-U) [above right = 7mm and 2mm of og] {$\Rightarrow$};
        \node (T-U-label) [above = -1mm of T-U] {T\textcolor{purple-cb}{$\bowtie$}U};
    \end{tikzpicture}
    \begin{subfigure}[t]{0.3\columnwidth}
        \centering
        \begin{tikzpicture}[
            v/.style={circle, draw=black, minimum size=26pt},
            e/.style={ultra thick}
        ]
            \node[v, node distance=2.8cm] (R2) [right = of R] {R};
            \node[v] (S2) [right=of R2] {S};
            \node[v, node distance=1.82cm] (TU2) [right=of U] {T\textcolor{purple-cb}{$\bowtie$}U};

            \draw[e, blue-cb] (R2) to (S2);
            \draw[e, green-cb] (R2) to (TU2);
            \draw[e, yellow-cb, bend left=5] (S2) to (TU2);
            \draw[e, orange-cb, bend right=5] (S2) to (TU2);
        \end{tikzpicture}
        \caption{JG~2}
        \label{fig:jg2}
    \end{subfigure}
    \begin{tikzpicture}[]
        \node (og) at (0,0) {};
        \node (R-S) [above right = 7mm and 2mm of og] {$\Rightarrow$};
        \node[node distance=-1mm] (R-S-label) [above = of R-S] {R\textcolor{blue-cb}{$\bowtie$}S};
    \end{tikzpicture}
    \begin{subfigure}[t]{0.1\columnwidth}
        \centering
        \begin{tikzpicture}[
            v/.style={circle, draw=black, minimum size=26pt},
            e/.style={ultra thick}
        ]
            \node[v, node distance=2.8cm] (RS3) [right = of R2] {R\textcolor{blue-cb}{$\bowtie$}S};
            \node[v] (TU3) [below=of RS3] {T\textcolor{purple-cb}{$\bowtie$}U};

            \draw[e, green-cb] (RS3) to (TU3);
            \draw[e, yellow-cb, bend left=10] (RS3) to (TU3);
            \draw[e, orange-cb, bend right=10] (RS3) to (TU3);
        \end{tikzpicture}
        \caption{JG~3}
        \label{fig:jg3}
    \end{subfigure}
    \caption{Transformation of a cyclic join graph into an acyclic one by folding vertices, i.e.\ computing join results of neighboring vertices.}
    \label{fig:folding}
\end{figure}
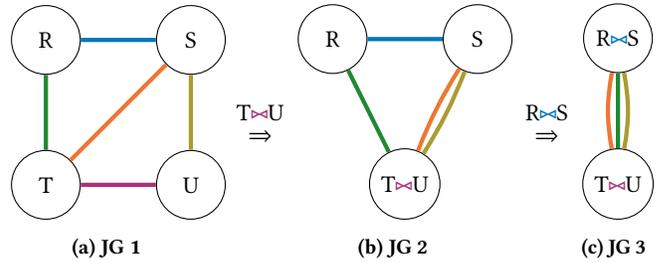
\begin{figure}[h]
    \centering
    \begin{subfigure}[t]{0.15\textwidth}
        \begin{tikzpicture}[
            v/.style={circle, draw=black, minimum size=26pt},
            e/.style={ultra thick}
        ]
                \Tree
                    [.\node (t1) {$\bowtie_{\textcolor{orange-cb}{\text{pred}} \land \textcolor{green-cb}{\text{pred}} \land \textcolor{yellow-cb}{\text{pred}}}$};
                        [.$\bowtie_{\textcolor{blue-cb}{\text{pred}}}$
                            [.R ]
                            [.S ]
                        ]
                        [.$\bowtie_{\textcolor{purple-cb}{\text{pred}}}$
                            [.T ]
                            [.U ]
                        ]
                    ]
        \end{tikzpicture}
        \caption{LQP for JG~1.}
        \label{fig:lqp-jg1}
    \end{subfigure}
    \begin{subfigure}[t]{0.15\textwidth}
        \begin{tikzpicture}[
            v/.style={circle, draw=black, minimum size=26pt},
            e/.style={ultra thick}
        ]
                \Tree
                    [.\node (t1) {$\bowtie_{\textcolor{orange-cb}{\text{pred}} \land \textcolor{green-cb}{\text{pred}} \land \textcolor{yellow-cb}{\text{pred}}}$};
                        [.$\bowtie_{\textcolor{blue-cb}{\text{pred}}}$
                            [.R ]
                            [.S ]
                        ]
                        [.\node {T\textcolor{purple-cb}{$\bowtie$}U}; ]
                    ]
        \end{tikzpicture}
        \caption{LQP for JG~2.}
        \label{fig:lqp-jg2}
    \end{subfigure}
    \begin{subfigure}[t]{0.15\textwidth}
        \begin{tikzpicture}[
            v/.style={circle, draw=black, minimum size=26pt},
            e/.style={ultra thick}
        ]
                \Tree
                    [.\node (t1) {$\bowtie_{\textcolor{orange-cb}{\text{pred}} \land \textcolor{green-cb}{\text{pred}} \land \textcolor{yellow-cb}{\text{pred}}}$};
                        [.\node {R\textcolor{blue-cb}{$\bowtie$}S}; ]
                        [.\node {T\textcolor{purple-cb}{$\bowtie$}U}; ]
                    ]
        \end{tikzpicture}
        \caption{LQP for JG~3.}
        \label{fig:lqp-jg3}
    \end{subfigure}
    \caption{A potential logical query plan (LQP) for each join graph given in ~\autoref{fig:folding}.}
    \label{fig:logical-plan}
\end{figure}
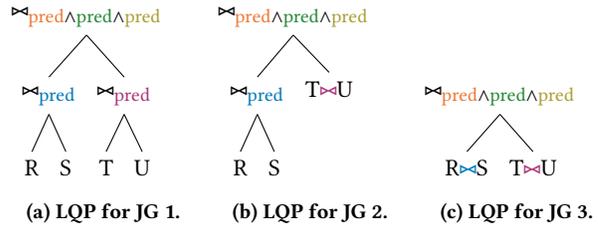

Each fold represents the temporary view of two or more joined relations (depending on how many relations are contained in a fold).
Therefore, this folding process is essentially analogous to what happens when successively computing a join result based on a specific join order.

Let's look at \autoref{fig:logical-plan} for example, which shows some potential logical query plans (LQP) for each join graph depicted in~\autoref{fig:folding}.
For the initial join graph (JG~1), there is no join order given, i.e.\ we are free to join whatever relations along the edges in the join graph.
In Figure~\ref{fig:lqp-jg1} we have a bushy tree representing the the join (R$\bowtie$S)$\bowtie$(T$\bowtie$U).
After we fold the first two relations, we basically substitute the rightmost subtree (joining T and U) with the newly computed join result in Figure~\ref{fig:lqp-jg2}.
Finally, we also replace the left subtree with the join result of R and S in Figure~\ref{fig:lqp-jg3}.
Note, that all three join graphs have multiple different valid logical join orders and we only depict a specific one.

Based on this analogy, the correctness of our algorithm follows naturally from the commutativity and associativity of joins.
Since the join graph does not define a join order, we can join any two relations $x$ and $y$.
Performing this join leaves the overall join result of the resulting join graph unchanged.
We basically just follow a join order where every join folds two nodes into a view, i.e.\ we obtain smaller join graph consisting of base relations and intermediate views/join results.
Obviously, we can repeat this process until the resulting join graph is acyclic and based on this, compute our semi-join reduction.

\subsection{Putting It All Together}
\label{subsec:putting-it-all-together}
Now that we have introduced the means to transform cyclic queries into tree queries, we can present our final \textsc{SELECT RESULTDB} \autoref{alg:select-resultdb}.
\begin{algorithm}[H]
    \centering
    \begin{algorithmic}[1]
        \Function{select\_resultdb}{$G$}
            \If{$G$.is\_cyclic()}
                \State $G$ = $G$.fold\_join\_graph($G$) \Comment{transform into acyclic jg}
            \EndIf
            \State $G$ = reduce\_relations($G$) \Comment{use Yannakakis' algorithm}

            \For{$r \in G.\text{relations}$}
                \If{$r$.is\_view()} \Comment{$r$ could be a join result}
                    \State $G$.split($r$) \Comment{split $r$ by projecting onto base relations}
                \EndIf
            \EndFor
            \State return $G$.relations \Comment{result database}
        \EndFunction
    \end{algorithmic}
    \caption{\textsc{SELECT RESULTDB} on arbitrary join graph topologies.}
    \label{alg:select-resultdb}
\end{algorithm}
Given an arbitrary join graph, we first check if the join graph is cyclic in line~2.
If that is the case, we use our folding algorithm in line~3 to construct an acyclic join graph.
Afterwards, we can reuse Yannakakis' algorithm to reduce the relations in line~4.
Note that at this point, a relation can also be a join result.
However, the algorithm just treats them as regular relations and performs the semi-joins based on the modified join predicates.
These join predicates can now be conjunctions of joins predicates (cf.~\autoref{fig:folding}).
Finally, we have to break up possible views/joins again in lines~5-7.
For this, we iterate over all nodes in the join graph (line~5) and check if this node is a view (line~6).
If so, we split up this node again (line~7) by projection onto each involved base relation (we have to store this information somewhere).
In general, this operation can basically be seen as the inverse of the folding algorithm.
Finally, we can return the now reduced relations of our join graph $G$.

%% file: related.tex

\section{Related Work}

\noindent\textbf{SQL Flaws.}
SQL having some flaws is no secret.
There are many publications~\cite{date_critique-sql, taipalus_errors-sql, grant_null-values, dittrich_gap-olap-sql} and blog posts~\cite{sql-flaws, we-can-do-better-than-sql} that present some major shortcomings of SQL as a language.
For example, they criticize SQL's NULL value semantic, that SQL is hard to compose due to its different kinds of expressions (table vs scalar), or that it is generally very inconsistent and error-prone.
In this work, we also discuss a number of problems which mainly come from the denormalization of tables through joins.
In addition, we provide a single keyword extension that is able to fix all problems mentioned in this work and even addresses some of the flaws mentioned in related work.

\noindent\textbf{SQL Extensions.}
Throughout the years, many SQL extensions have been introduced.
These extensions come in many different variations.
Regarding single keyword extensions, the Data Cube~\cite{gray_data-cube} operator and the Skyline~\cite{borzsonyi_skyline} operator are probably the most well-known.
Both operators add a very specific new functionality to SQL.
For example, the Data Cube operator enables N-dimensional aggregate computation while the Skyline operator allows for filtering out ``interesting'' data points.
Our work also introduces a new keyword that provides specific functionality, namely the computation of a result database instead of a single-table result.
To the best of our knowledge, this is the first work with this contribution.

Another way of extending SQL is to move away from the traditional relational data model and introducing a query language tailored to semi-structured or unstructured data.
SQL++~\cite{ong_sql++} is a semi-structured query language of AsterixDB~\cite{asterixdb} that basically represents a superset of the SQL and JSON data model.
Extending SQL with JSON features enables the arbitrary nesting and composition of data values.
This essentially provides us with an additional option to retrieve data in a way that lets us populate the UI-component in~\autoref{fig:nested}, our motivating problem.
Note that this is basically a more sophisticated approach to using SQL's native array-logic to return nested tables, similar to Option~2 discussed in \autoref{sec:introduction}.
While this is definitely a valid approach, our proposed extension explicitly stays in the relational word, keeping schema and relational information.
In addition, it is minimally invasive since it only requires the addition of a single keyword instead of using a new query language with many different features.

\noindent\textbf{Semi-joins.}
To implement our idea to return a result database, the fundamental idea is to use semi-joins to reduce all involved relations.
Therefore, the concept of using semi-joins is a central part of this work.

Bernstein et. al~\cite{bernstein_semi-joins} already introduces the idea to use semi-joins to solve relational queries.
By ``solving a relational query'' they refer to efficiently computing the reduced relations using semi-joins and subsequently computing the final join result using these reduced relations.
This algorithm is also well-known as Yannakakis'~\cite{yannakakis_acyclic-database-schemes} algorithm.
In their work, they show that queries with a tree-structured query graph can be solved using semi-joins.
Moreover, they show that queries with cyclic query graph can generally either not be solved at all or only using large \emph{semi-join programs}.
Both works were an inspiration for our algorithmic solution.

\noindent\textbf{Sideways Information Passing \& Factorization.}
The underlying idea of sideways information passing (SIP) is to optimize query processing by exchanging information between arbitrary parts in a query plan.
This optimization is often aimed at reducing intermediate (join) results by \emph{reducing} relations early on.
Therefore, a semi-join reduction~\cite{bernstein_semi-joins} is actually a special case of SIP.
The application of the concept of SIP is very broad.
In~\cite{shrinivas_vertica}, Shrinivas et al. present how sideways information passing in the context of materialization strategies can be used to improve the performance in the Vertica Analytical Database~\cite{lamb_vertica-database}.
Neumann and Weikum~\cite{neumann_RDF-graphs}, show how SIP can be used to speed up index scans at query runtime in RDF graphs.
Another work~\cite{zhu_looking-ahead} by Zhu et al. is able to produce robust query plans in star schemas by making use of bloom filters, a SIP data structure.
While SIP applies a certain reduction in a specific scenario, we unconditionally reduce all relations that are part of a query.
Furthermore, we explicitly want to compute this reduced state of a relation whereas SIP only uses it as a means to speed up query processing.

Factorized Databases~\cite{olteanu_factorized} is another concept that tries to minimize intermediate join results.
However, in contrast to the aforementioned SIP methods, factorization is essentially a compression techniques for join results that tries to get rid of redundancies.
With that, it also targets the problem of relational information redundancies (cf. Problem 1 in \autoref{sec:introduction}).
In contrast to factorized databases, we primarily try to avoid the redundancies in join results instead of hiding those redundancies using compression techniques.

%% file: eval.tex

\section{Experiments}
In this section, we are going to have a closer look at the rewrite methods discussed in~\autoref{sec:rewrite}.
In particular, we will evaluate the end-to-end client-side runtime of the presented methods.
As a baseline, we use the traditional single-table query performance.
In that respect, we will also briefly compare the query result set sizes of traditional single-table queries with \textsc{SELECT RESULTDB}.

\emph{Please note: In the current state of research, we are in the process of integrating the algorithm discussed in \autoref{sec:algorithm} directly into a database management system. This part of the evaluation will be added in a later version of this work.}

\noindent\textbf{Setup.} The experiments are conducted on the TPC-H benchmark\footnote{\url{https://www.tpc.org/tpch/}} with scale factor 1.
In addition, we use a PostgreSQL\footnote{\url{https://www.postgresql.org/}} database with version 15.4-2 and default settings.
Further, we measure the client-side runtime using the interactive PostgreSQL terminal \emph{psql}\footnote{\url{https://www.postgresql.org/docs/current/app-psql.html}}
and the \texttt{timing on/off} commands.
With the client-side runtime we explicitly include not only the query optimization and processing time but also the transmission time to the client.
The underlying operating system is Arch Linux with kernel version 6.6.1.

\subsection{End-To-End Client-Side Runtime}
The performance of our rewrite methods is fundamentally dependent on two factors.
First, the complexity and size of the join and second, how many different relations are part of the query result.

The one thing all rewrite methods have in common is that they execute one query for each unique relation in the \textsc{SELECT} clause.
Take the query based on the TPC-H schema in \autoref{lst:tpc-h-query} as an example.
This query just joins five different relations along their foreign-key relationships and has a filter on one of the relations.
This filter \texttt{ps\_supplycost < \{sup\_cost\}} lets us control the overall query result size.
Since only two relations, namely \emph{part} and \emph{supplier}, are part of the single-table query result, we need to perform exactly two logical semi-joins.
\begin{lstlisting}[language=sql, caption={Query on TPC-H dataset joining five tables along the foreign-key relationships with one additional filter predicate.}, label={lst:tpc-h-query}]
select
    -- post-join:
    -- p_partkey,
    -- ps_partkey, ps_suppkey,
    -- s_suppkey,
    p_name,
    p_brand,
    s_name
from
    supplier,
    nation,
    region,
    partsupp,
    part
where
    ps_supplycost < {sup_cost}
    and r_regionkey = n_regionkey
    and n_nationkey = s_nationkey
    and s_suppkey = ps_suppkey
    and ps_partkey = p_partkey;
\end{lstlisting}

\begin{figure*}[h!]
    \begin{subfigure}[h]{0.48\textwidth}
        \centering
        \includegraphics[keepaspectratio, width=\textwidth]{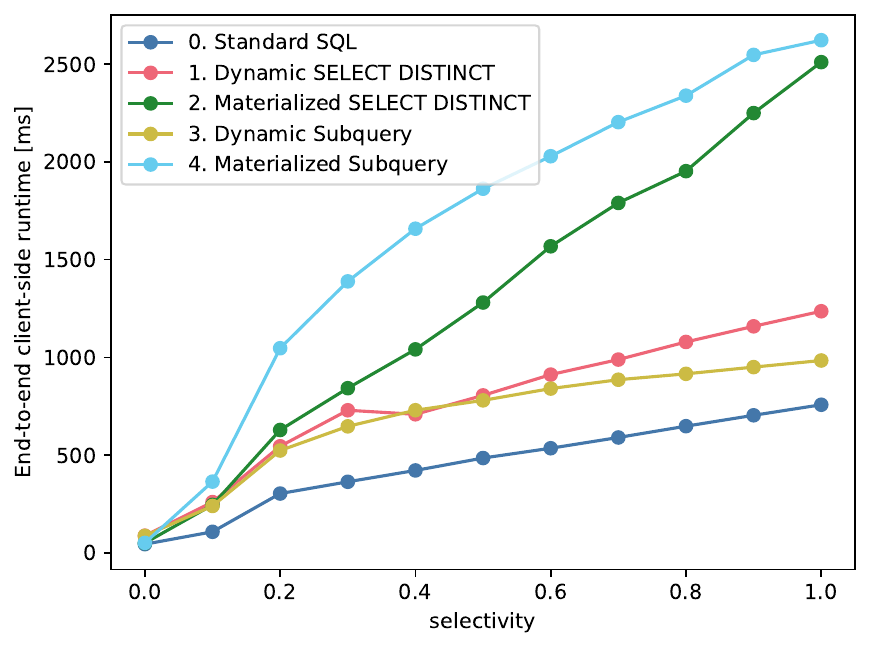}
        \caption{Client-side runtime.}
        \label{subfig:runtime}
    \end{subfigure}
    \hfill
    \begin{subfigure}[h]{0.48\textwidth}
        \centering
        \includegraphics[keepaspectratio, width=\textwidth]{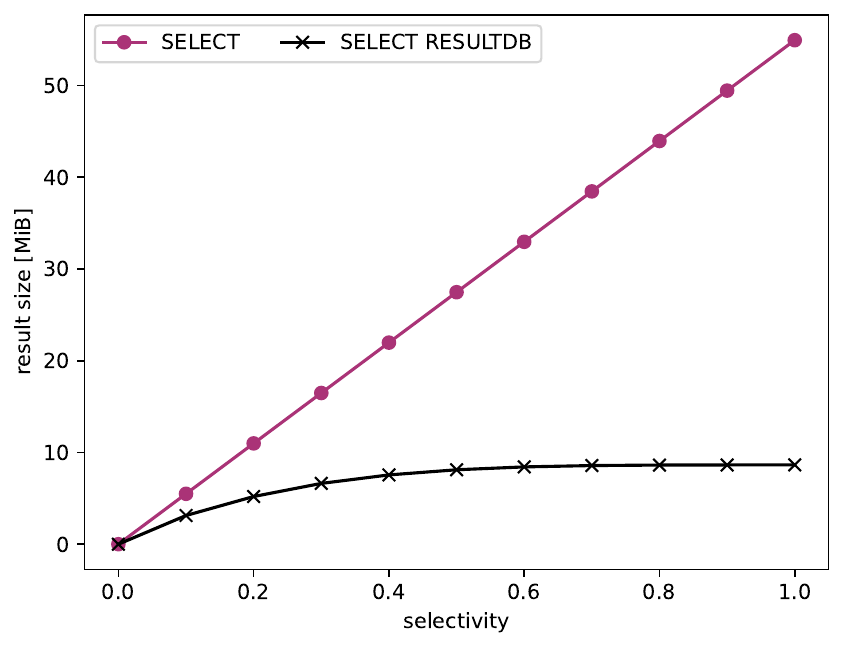}
        \caption{Query result size.}
        \label{subfig:memory-consumption}
    \end{subfigure}
    \caption{Performance evaluation of the query given in \autoref{lst:tpc-h-query} for different selectivities.}
    \label{fig:tpc-h-aq1}
\end{figure*}

Figure~\ref{subfig:runtime} shows the end-to-end client-side runtime for different selectivity values.

Overall, we can observe that the standard SQL query that produces a single-table result scales linearly with the selectivity factor and is slightly faster than some of the rewrites methods.
For a selectivity factor of 1, standard SQL has a runtime of about 750~ms. In comparison, the fastest rewrite method (Dynamic Subquery) is roughly 200ms slower.
However, note that the standard SQL query and the rewrite methods should be compared with caution as both result sets differ per definition.

Rewrite~Method~1 (RWM~1: Dynamic \textsc{SELECT DISTINCT}) and Rewrite~Method~3 (RWM~3: Dynamic Subquery) have comparable runtimes.
For larger selectivities, the gap of RWM~3 to RWM~1 gets a little larger.
When comparing the physical execution plans (created using the \textsc{EXPLAIN ANALYZE} keyword of PostgreSQL) of both methods, we can see that in RWM~3 \emph{Hash Semi Join}s are used.
Therefore, the better performance of RWM~3 can probably be attributed to the usage of semi-joins instead of inner joins.

The other two rewrite methods, Rewrite~Method~2 (RWM~2: Materialized \textsc{SELECT DISTINCT}) and Rewrite~Method~4 (RWM~4: Materialized Subquery), are rather slow in comparison.
The most obvious reason for this is the relatively large join result that has to be materialized temporarily.
Apparently, the tradeoff between the upfront materialization costs versus the repeated join computation costs does not pay off for this specific query.
Having said that, these types of methods could obviously still be beneficial if we access more than two different relations.
Comparing RWM~2 and RWM~4, we can see that the method with the \textsc{SELECT DISTINCT} performs for most selectivities considerably better.
Again looking at the physical execution plan, we can see that for RWM~2 we only execute an aggregation on the materialized view (MV).
In comparison, RWM~4 executes one inner join and one semi-join for both subqueries.
We believe this additional computation to be more costly than the aggregation.

To summarize, even though the the rewrite methods perform slightly worse compared to the standard SQL query, this difference is not huge.
In addition, we avoid all the problems (1-7) discussed in \autoref{sec:introduction}.
Moreover, we believe that in many cases where we deviate from foreign-key joins with one-to-one or one-to-many relationships, the potential gain of our approach is high.

\subsection{Memory Consumption}
Finally, we briefly touch on the result set sizes of normal SQL versus our approach.
Figure~\ref{subfig:memory-consumption} shows the query result size for the standard SQL query (\textsc{SELECT}) and our result database (\text{SELECT RESULTDB}) for the different selectivity values.
As expected, the output size of the single-table join result scales linearly with the selectivity.
In contrast, we can see that the combined size of the two individual relations, which are part of the result database, approaches the maximal size of around 8.5~MiB very fast.
Furthermore, this size is considerably smaller than the single-table size which goes up to 55~MiB for a selectivity factor of 1.
This clearly shows that the final result of the single-table query contains many redundancies which we are able to prune.

%% file: conclusion.tex

\section{Conclusion}
SQL comes with the very hard limitation that each query result is shoehorned into a single table.
In this work, we initially discuss the unexpected long list of different problems that stem from this limitation.
To address these problems, we propose to extend the SQL \textsc{SELECT} clause by a single keyword: \textsc{RESULTDB}.
This extension lets us return a subdatabase, i.e.\ a subset of the relations as well as a subset of the tuples of those relations, instead of a single, possibly denormalized, relation.
Furthermore, we introduce a formalization of our SQL extension showing that it is well-defined and has clear semantics.

We present two different approaches to support our new functionality.
First, we propose four SQL-based rewrite methods that allow us to transform traditional SQL queries into semantically equivalent queries returning a result database.
This allows for very easy integration, especially in closed source database systems.
Second, we propose an efficient algorithm that can be integrated directly into a DBMS.
We also show some promising initial experimental results for our rewrite methods regarding client-side runtime and join result size.

\noindent\textbf{Future Work.}
Our primary objective for the moment is to complete the integration of our proposed algorithm into a state-of-the-art database management system.
Afterwards, we want to extend our initial experiments. In particular, we aim to more thoroughly evaluate our rewrite methods and benchmark our algorithm.

Another very interesting area is query optimization in the context of semi-join reductions.
Yannakakis' algorithm, as presented in this work, is rather vague.
We want to look in more detail into how we choose for example the root node or in which order we apply the semi-joins.
The same applies to our folding algorithm which exhibits some optimization potential.

Furthermore, our current work focuses purely on the \emph{data retrieval} aspect of SQL, i.e.\ SPJ queries.
However, we also have ongoing research regarding \emph{data transformation}.
We envision to do data transformation alongside data retrieval, i.e.\ strictly separating both aspects, in contrast to SQL that intransparently mixes them.
Therefore, we want to intensify and publish our work regarding data transformation.